\documentclass[prl,twocolumn,superscriptaddress,floatfix,longbibliography]{revtex4}
            \usepackage{enumerate}
            \usepackage{graphicx}
            \usepackage{amsmath}
            \usepackage{mathtools}
            \usepackage{xcolor}
            \usepackage{amsfonts}
            \usepackage{amssymb}
            \usepackage{bm}
            \usepackage{units}
            \usepackage[normalem]{ulem}
            \newcommand{\ket}[1]{| #1 \rangle}
            \newcommand{\bra}[1]{\langle #1 |}

\begin{document}
            \pdfoutput=1

            \title{Magnon Shake-up: Entanglement Generation and Sensing}
            \author{Vahid Azimi-Mousolou\footnote{Electronic address: \\ vahid.azimi-mousolou@physics.uu.se\\ v.azimi@sci.ui.ac.ir}}
            \affiliation{Department of Applied Mathematics and Computer Science, 
            Faculty of Mathematics and Statistics, 
            University of Isfahan, Isfahan 81746-73441, Iran}
            \affiliation{Department of Physics and Astronomy, Uppsala University, Box 516, 
            SE-751 20 Uppsala, Sweden}

             \author{Erik Sj\"oqvist}
            \affiliation{Department of Physics and Astronomy, Uppsala University, Box 516, 
            SE-751 20 Uppsala, Sweden}

             \author{Anna Delin}
            \affiliation{Department of Applied Physics, School of Engineering Sciences, KTH Royal Institute of Technology, AlbaNova University Center, SE-10691 Stockholm, Sweden}
            \affiliation{Swedish e-Science Research Center, KTH Royal Institute of Technology, SE-10044 Stockholm, Sweden}
            \affiliation{Wallenberg Initiative Materials Science for Sustainability (WISE), KTH Royal Institute of Technology, SE-10044 Stockholm, Sweden}
            
            \author{Olle Eriksson}
            \affiliation{Department of Physics and Astronomy, Uppsala University, Box 516, 
            SE-751 20 Uppsala, Sweden} 
            \affiliation{WISE-Wallenberg Inititative Materials Science, Uppsala University, Box 516, SE-751 20 Uppsala, Sweden}
            
            \date{\today}
            
            \begin{abstract}
            Shake-up phenomena, rooted in the sudden approximation and many-body quantum dynamics, unveil critical characteristics of quantum systems, with wide-ranging applications in molecular spectroscopy and electronic structure studies. Here, this principle is extended to magnetic systems, and magnon shake-up structures and their relevance in generating and sensing magnon-magnon entanglement are demonstrated. This not only reveals new insights into shake-up phenomena and quantum magnetism but also paves the way for new opportunities in quantum technologies.
            \end{abstract}
            \maketitle
            
            The fusion of ultrafast laser technology with spin dynamics has generated immense interest in photomagnonics, a field poised to advance magnon-based quantum information processing 
            \cite{ Beaurepaire1996, van-Kampen2002, Kimel2004, Kimel2005, Satoh2010}. 
            Recent observations of subpicosecond demagnetization using femtosecond laser pulses have unveiled new possibilities for manipulating spin waves with exceptional precision.
            This development has catalyzed the emergence of quantum magnonics \cite{Yuan2022}, where ultrafast lasers can be utilized to create and control entangled magnon states \cite{Bossini2019, hales2023}. These states are crucial for both rapid magnetization control and advancements in quantum information science. By modifying exchange interactions with strong laser pulses, two-mode entangled magnon states can be generated, bridging ultrafast dynamics and quantum technologies. Notably, such rapid perturbations of the Hamiltonian may introduce a novel quantum magnonics phenomenon analogous to shake-up structures observed in molecular spectroscopy and electronic structure studies \cite{Manne1970, Tillborg1993, Doniach1970, Rosencwaig1971, Carlsson1974}.
            Here, we demonstrate magnonics shake-up structures and leverage them to generate and detect magnon-magnon entanglement, enabling new possibilities for high-speed data processing and quantum computing.
            
            The shake-up structure has been studied theoretically in X-ray photoelectron spectroscopy \cite{Manne1970, Tillborg1993} and observed experimentally in works such as \cite{Doniach1970, Rosencwaig1971, Carlsson1974}. The core concept can be understood in terms of sudden perturbations to the system, such as those caused by X-ray radiation or ultrafast laser-driven excitation.
            To adapt the shake-up mechanism to the magnonics regime, we consider a magnonics system described by the Hamiltonian $H$, initially prepared in the $N$-magnon state, $\ket{\psi_{N}} = \ket{N}$. 
            We assume the system undergoes an abrupt perturbation, such as a light-induced change in exchange interaction, on a very short timescale $\tau$, resulting in a new magnonics system described by the Hamiltonian $H(\tau)$. The process is so rapid that the system cannot adjust and thus remains in its initial state $\ket{N}$ according to the sudden approximation  \cite{Messiah1962}. However, this state is no longer a pure eigenstate of the instantaneous Hamiltonian $H(\tau)$. Instead, it becomes a mixture of eigenstates of $H(\tau)$ described by
            \begin{eqnarray}
            \ket{\psi_{N}} = \sum_{n=0}^{\infty} P(\tau, n)\ket{\Psi_{n}(\tau)} =\ket{\psi_{R}}+\ket{\psi_{E}},
            \label{SHUP}
            \end{eqnarray}
            where $\ket{\Psi_{n}} (\tau)$ represents the $n$-magnon eigenstate of the modified Hamiltonian $H(\tau)$, and $P(\tau, n) = \langle\Psi_{n}(\tau)|\psi_{N}\rangle$ describes the transition probability $|P(\tau, n)|^{2}$ to the $n$-magnon state. 
            
            The physical situation can be described as follows: upon excitation, e.g., by absorption of light in an ultrafast pump-probe experiment, a portion of the quantum information stored in the $N$-magnon state of the system is expelled. At the measurement time $\tau$, the state of the system can be decomposed into a superposition of two components, as shown in Eq.\ \eqref{SHUP}. The first component, $\ket{\psi_{R}} = P(\tau, N)\ket{\Psi_{N}(\tau)}$, represents the remaining information of the system within the $N$-magnon state $\ket{N}$. The second component, $\ket{\psi_{E}} = \sum_{n=0, n \neq N}^{\infty} P(\tau, n)\ket{\Psi_{n}(\tau)}$, corresponds to the expelled portion of the information, referred to as the free information. This expelled portion may induce further excitation in the rest of the magnon system. Following an ultrafast excitation, and relying on the approximation that the magnon state $\ket{N}$ does not have time to change, the energy distribution in the magnon system can be expressed as
      \begin{eqnarray}
            \bra{\psi_{N}}H(\tau)\ket{\psi_{N}}&=&\bra{\psi_{R}}H(\tau)\ket{\psi_{R}}+\bra{\psi_{E}}H(\tau)\ket{\psi_{E}}\nonumber\\
            &=&|P(\tau, N)|^{2}\varepsilon_{N}+ \sum_{{\substack{n=0 \\ n \neq N}}}^{\infty}|P(\tau, n)|^{2}\varepsilon_{n},\ \ \ \ \ 
            \label{SHUP2}
      \end{eqnarray}     
where  $\varepsilon_{n}=\bra{\Psi_{n}(\tau)}H(\tau)\ket{\Psi_{n}(\tau)}$. Note that in this expression 
$\varepsilon_{N}$ is not necessarily the same as the energy of the N magnon state of the unperturbed Hamiltonian, since $\bra{\Psi_{N}}H(\tau)\ket{\Psi_{N}} \neq \bra{\Psi_{N}}H\ket{\Psi_{N}}$.
The second summation in Eq.~\eqref{SHUP2} corresponds to the energy distribution from the expelled state, leading to the occurrence of additional peaks in the energy measurement of the magnon system. We refer to these peaks as magnonics shake-up peaks. 

The magnon shake-up described by Eq.~\eqref{SHUP2} is analogous to the shake-up structure observed in photoelectron spectroscopy \cite{Manne1970, Tillborg1993}, 
and it can in a similar fashion be experimentally verified through the associated energy shift, as well as modification of the intensity of the magnon excitation, expressed by the probability $|P(\tau, N)|^{2}$. Both entities are detectable experimentally, e.g., by the dynamical structure factor and below we will illustrate with concrete examples how the terms in Eq.~\eqref{SHUP2} become visible in magnon excitations. 
            
To demonstrate the magnon shake-up process in detail, we explore its role in generating and sensing quantum entanglement among magnons in magnetic spin systems. 
We consider an antiferromagnet with easy-axis anisotropy, that is subjected to an external magnetic field. This is described by the spin Hamiltonian 
        \begin{eqnarray}
            H=J\sum_{\langle i, j\rangle}\mathbf{S}_{i}\cdot\mathbf{S}_{j}-\mathcal{K}\sum_{i}\left(S_{i}^{z}\right)^{2}-\gamma\hbar\sum_{i}\mathbf{B}\cdot\mathbf{S}_{i} 
            \label{BSM}
        \end{eqnarray}
with nearest neighbor antiferromagnetic exchange $J>0$, anisotropy $\mathcal{K}$ provides an easy axis along the $z$-axis, and the external magnetic field is $\mathbf{B}=B\mathbf{e}_z$.
Here, $\gamma=g\mu_{B}/\hbar$ is the gyromagnetic ratio with $g$ being the spectroscopic splitting factor, $\mu_{B}$ the Bohr magneton, and $\hbar$ the reduced Planck constant. This model describes a wide class of magnetic materials, including hexagonal SrMnO$_3$ (space group P6$_3$/mmc), LaFeO$_3$ (space group Pnma), MnF$_{2}$ and FeF$_{2}$ (both in space group P4$_2$/mnm).
            
One may identify two spin sublattices and use the linear Holstein-Primakoff (HP) transformation,
$S_{i}^{z} = S-a_{i}^{\dagger} a_{i}$, $S_{j}^{z} = b_{j}^{\dagger} b_{j}-S$, $S_{i}^{+} = \sqrt{2S} a_{i}$, $S_{j}^{+} = \sqrt{2S} b_{j}^{\dagger}$
where $S_{i,j}^{\pm} = S_{i, j}^{x} \pm i S_{i, j}^{y}$ and $S$ is the spin magnitude. Each of the bosonic operators $a_{i}^{\dagger}$ and $b_{j}^{\dagger}$ ($a_{i}$ and $b_{j}$) creates (annihilates) an HP boson, representing spin excitations at lattice sites $i$ and $j$ in the corresponding sublattices. By applying the HP transformation followed by a Fourier transformation,
        $a_{i} = \sqrt{\frac{2}{N}} \sum_{\mathbf{k}} e^{-i \mathbf{k} \cdot \mathbf{r}_{i}} a_{\mathbf{k}}$ and $b_{j} = \sqrt{\frac{2}{N}} \sum_{\mathbf{k}'} e^{-i \mathbf{k}' \cdot \mathbf{r}_{j}} b_{\mathbf{k}'}$,
the system Hamiltonian in Eq.~\eqref{BSM} can be expressed in terms of bosonic operators in momentum $\mathbf{k}$-space as \cite{note}
        \begin{eqnarray}
            H_{\mathbf{k}}= \omega_{a}a_{\mathbf{k}}^{\dagger}a_{\mathbf{k}}+\omega_{b}b_{-\mathbf{k}}^{\dagger}b_{-\mathbf{k}}+g_{\mathbf{k}}a_{\mathbf{k}}b_{-\mathbf{k}}+g^{*}_{\mathbf{k}}a_{\mathbf{k}}^{\dagger}b_{-\mathbf{k}}^{\dagger},
            \label{eq:FBT}
        \end{eqnarray}
        with  $\omega_{a}=\omega+B$ and $\omega_{b}=\omega-B$, where 
        \begin{eqnarray}
            \omega=S(ZJ+2\mathcal{K}),\ \ \ \ \ 
            g_{\mathbf{k}}=SZJ\gamma_{\mathbf{k}}.
            \label{eq:FBTC}
        \end{eqnarray}
Here $\gamma_{\mathbf{k}}=\frac{1}{Z}\sum_{\bm{\delta}}e^{i\mathbf{k}\cdot\bm{\delta}}$ is the lattice geometric parameter, with $Z$ being the coordination number identifying the number of nearest neighbors for each site in the spin lattice, and $\bm{\delta}$ being a vector connecting each site to its neighboring sites.
                       
Under $SU(1,1)$ Bogoliubov transformation, $a_{\mathbf{k}}=\eta_{\mathbf{k}}\alpha_{\mathbf{k}}+\zeta_{\mathbf{k}}\beta_{-\mathbf{k}}^{\dagger}$ and
$b_{-\mathbf{k}}^{\dagger}=\zeta^{*}_{\mathbf{k}}\alpha_{\mathbf{k}}+\eta_{\mathbf{k}}\beta_{-\mathbf{k}}^{\dagger}$,
where $\eta_{\mathbf{k}} =\cosh(r_{\mathbf{k}})$ and $\zeta_{\mathbf{k}} = e^{i\phi_{\mathbf{k}}}\sinh(r_{\mathbf{k}})$ are determined by finite-valued parameters
            \begin{eqnarray}
            \tanh r_{\mathbf{k}}=\frac{1-\sqrt{1-|\Gamma_{\mathbf{k}}|^{2}}}{|\Gamma_{\mathbf{k}}|}\ge 0,\ \ \ \ \ \nonumber\\
            \Gamma_{\mathbf{k}}=\frac{g_{\mathbf{k}}}{\omega},\ \ \ \ \phi_{\mathbf{k}}=\pi-\arg[\Gamma_{\mathbf{k}}],
            \label{SQPP}
            \end{eqnarray}
we obtain the following diagonal Hamiltonian
             \begin{eqnarray}
            H_{\mathbf{k}} = 
            \epsilon_{\alpha_{\mathbf{k}}}\alpha_{\mathbf{k}}^{\dagger}\alpha_{\mathbf{k}} + \epsilon_{\beta_{-\mathbf{k}}}\beta_{-\mathbf{k}}^{\dagger}\beta_{-\mathbf{k}},
             \label{DHH}
            \end{eqnarray}
            in which the magnon dispersion relations are $\epsilon_{\alpha_{\mathbf{k}}}=\epsilon_{\mathbf{k}}+B$ and $\epsilon_{\beta_{-\mathbf{k}}}=\epsilon_{\mathbf{k}}-B$ with 
            \begin{eqnarray}
            \epsilon_{\mathbf{k}}&=&\omega\left[\cosh(2 r_{\mathbf{k}})-|\Gamma_{\mathbf{k}}|\sinh(2 r_{\mathbf{k}})\right].
            \end{eqnarray}
            
Equation \eqref{DHH} implies the energy eigenstates: 
            \begin{eqnarray}
            \ket{\psi_{mn}^{(\mathbf{k})}}&=&\ket{m; \alpha_{\mathbf{k}}}\ket{n;\beta_{-\mathbf{k}}}\nonumber\\
            &=&\frac{1}{\sqrt{m!n!}}[\beta_{-\mathbf{k}}^{\dagger}]^{n}[\alpha_{\mathbf{k}}^{\dagger}]^{m}\ket{0; \alpha_{\mathbf{k}}}\ket{0; \beta_{-\mathbf{k}}},
            \label{SQEES}
            \end{eqnarray}
which describe separable twin magnon modes, $\alpha_{\mathbf{k}}$ and $\beta_{-\mathbf{k}}$, propagating in opposite directions. In particular, the state $\ket{\psi_{00}^{(\mathbf{k})}} = \ket{0; \alpha_{\mathbf{k}}} \ket{0; \beta_{-\mathbf{k}}}$ represents the separable two-mode vacuum state. For given values of $m$ and $n$, the state in Eq.~\eqref{SQEES} specifies $\ket{\psi_{N}}$, i.e., $\ket{\psi_{N}} = \ket{\psi_{mn}^{(\mathbf{k})}}$ with $N = m + n$.
              
Given the two-mode magnon system in 
Eq.\,\eqref{DHH}, the magnon shake-up structure can be demonstrated by an abrupt change in the exchange parameter $J$.
Light-induced perturbations to the exchange interaction have been studied and derived, e.g., from the electronic Hubbard model, resulting in a modification of the Hamiltonian in Eq.\ \eqref{BSM} by an amount \cite{Bossini2019, Mentink2015, Mikhaylovskiy2015} 
            \begin{eqnarray}
            \delta H =f(\tau)\sum_{i, \bm{\delta}}\Delta J(\bm{\delta})\mathbf{S}_i \cdot \mathbf{S}_{i+\bm{\delta}},
            \label{eq:pulse}
            \end{eqnarray}
where $f(\tau)$ represents a normalized envelope function describing the time profile of the excitation pulse. 
For a bond along a given direction $\bm{\delta}$, the light-induced modification of the exchange interaction takes the form
\begin{eqnarray}
\Delta J(\delta) &=& \frac{t^2(e\bm{\delta}\cdot \bm{E})^2}{4U(U^2-\hbar^2\omega^2)} = \frac{(\bm{\delta}\cdot \bm{E})^2}{a^2E^2} \Delta J, \nonumber\\
\Delta J &=& \frac{t^2e^2a^2E^2}{4U(U^2-\hbar^2\omega^2)},
\label{eq:DeltaJ}
\end{eqnarray}
in the approximation of super-exchange mechanism. Here, $t$ is the electronic hopping integral, $U$ is the onsite Coulomb interaction between electrons, $e$ is the elementary charge. The symbol $\omega$ represents the angular frequency of the optical electric field.  
The term $\bm{\delta} \cdot \bm{E} / (a E)$ denotes the projection of the optical electric field along the nearest-neighbor bond between two spins, given the lattice constant $a$ and the electric field amplitude $E$.
      
Following the bosonic transformations applied above, the time-dependent total Hamiltonian reads
            \begin{eqnarray}
            H_{\mathbf{k}}(\tau) &=&H_{\mathbf{k}}+\delta H_{\mathbf{k}}\nonumber\\
            &=&\epsilon_{\alpha_{\mathbf{k}}}(\tau)\alpha_{\mathbf{k}}^{\dagger}\alpha_{\mathbf{k}} +\epsilon_{\beta_{-\mathbf{k}}}(\tau)\beta_{-\mathbf{k}}^{\dagger}\beta_{-\mathbf{k}}+\nonumber\\
            &&\chi_{\mathbf{k}}(\tau)\alpha_{\mathbf{k}}\beta_{-\mathbf{k}}+\chi^{*}_{\mathbf{k}}(\tau)\alpha^{\dagger}_{\mathbf{k}}\beta^{\dagger}_{-\mathbf{k}},
            \label{TDTH}
            \end{eqnarray}
with $\epsilon_{\alpha_{\mathbf{k}}}(\tau)=\epsilon_{\mathbf{k}}(\tau)+B$ and $\epsilon_{\beta_{-\mathbf{k}}}(\tau)=\epsilon_{\mathbf{k}}(\tau)-B$.
Given $\Omega(\tau)=SZ\Delta Jf(\tau)$, we obtain 
 $\epsilon_{\mathbf{k}}(\tau)=\epsilon_{\mathbf{k}}+\delta\epsilon_{\mathbf{k}}(\tau)$, where 
 \begin{eqnarray}
           \delta\epsilon_{\mathbf{k}}(\tau)&=&\Omega(\tau)\left[\xi_{0}\cosh 2r_{\mathbf{k}}+|\xi_{\mathbf{k}}|\sinh 2r_{\mathbf{k}}\cos(\phi_{\mathbf{k}}+\varphi_{\mathbf{k}})\right],\nonumber\\
            \chi_{\mathbf{k}}(\tau)&=&\Omega(\tau)e^{-i\phi_{\mathbf{k}}}\left[\xi_{0}\sinh 2r_{\mathbf{k}}+\right.\nonumber\\
            &&\left. |\xi_{\mathbf{k}}|\left(\cosh 2r_{\mathbf{k}}\cos(\phi_{\mathbf{k}}+\varphi_{\mathbf{k}})+i\sin(\phi_{\mathbf{k}}+\varphi_{\mathbf{k}})\right)\right],\nonumber\\
            \xi_{\mathbf{k}}&=&\frac{1}{Za^2E^2}\sum_{\bm{\delta}}(\bm{\delta}\cdot \bm{E})^2e^{i\mathbf{k}\cdot\bm{\delta}}=|\xi_{\mathbf{k}}|e^{i\varphi_{\mathbf{k}}}.
\end{eqnarray}
            
At each instant $\tau > 0$, the Hamiltonian $H_{\mathbf{k}}(\tau)$ in Eq.~\eqref{TDTH} retains the same structure as the initial Hamiltonian $H_{\mathbf{k}}(0) = H_{\mathbf{k}}$ in Eq.~\eqref{eq:FBT}. Thus, following a similar procedure, we obtain the diagonal form of the instantaneous Hamiltonian as
             \begin{eqnarray}
            H_{\mathbf{k}}(\tau)  = 
            \varepsilon_{\alpha_{\mathbf{k}}}(\tau)\alpha_{\mathbf{k}}^{\dagger}(\tau)\alpha_{\mathbf{k}}(\tau) + \varepsilon_{\beta_{-\mathbf{k}}}(\tau)\beta_{-\mathbf{k}}^{\dagger}(\tau)\beta_{-\mathbf{k}}(\tau)\ \ \ 
            \label{IH}
            \end{eqnarray}
with instantaneous magnon dispersion relations $\varepsilon_{\alpha_{\mathbf{k}}}(\tau)=\varepsilon_{\mathbf{k}}(\tau)+B$ and $\varepsilon_{\beta_{-\mathbf{k}}}(\tau)=\varepsilon_{\mathbf{k}}(\tau)-B$ 
specified by    
            \begin{eqnarray}
            \varepsilon_{\mathbf{k}}(\tau)&=&\epsilon_{\mathbf{k}}(\tau)\left[\cosh(2 \Theta_{\mathbf{k}}(\tau))-| \Upsilon_{\mathbf{k}}(\tau)|\sinh(2 \Theta_{\mathbf{k}}(\tau))\right].\nonumber\\
            \label{LRDR}
            \end{eqnarray}
By assuming $(\alpha_{\mathbf{k}}(0), \beta_{-\mathbf{k}}(0))=(\alpha_{\mathbf{k}}, \beta_{-\mathbf{k}})$,  Eqs. \eqref{IH} and \eqref{LRDR} follow for $SU(1,1)$ Bogoliubov transformation
             \begin{eqnarray}
            \left(
            \begin{array}{cc}
              \alpha_{\mathbf{k}}    \\
               \beta_{-\mathbf{k}}^{\dagger}     
            \end{array}
            \right)=\left(
            \begin{array}{cc}
             u_{\mathbf{k}}(\tau)&  v_{\mathbf{k}}(\tau)    \\
            v^{*}_{\mathbf{k}}(\tau)&  u_{\mathbf{k}}(\tau)      
            \end{array}
            \right)\left(
            \begin{array}{cc}
              \alpha_{\mathbf{k}}(\tau)    \\
              \beta_{-\mathbf{k}}^{\dagger}(\tau)       
            \end{array}
            \right),
            \label{eq:BTfinal}
            \end{eqnarray}
where $u_{\mathbf{k}}(\tau) =\cosh(\Theta_{\mathbf{k}}(\tau))$ and $v_{\mathbf{k}}(\tau) =e^{i\Phi_{\mathbf{k}}(\tau)} \sinh(\Theta_{\mathbf{k}}(\tau))$ are obtained through  
            \begin{eqnarray}
            \tanh\Theta_{\mathbf{k}}(\tau)=\frac{1-\sqrt{1-| \Upsilon_{\mathbf{k}}(\tau)|^{2}}}{|\Upsilon_{\mathbf{k}}(\tau)|}\ge 0,\nonumber\\
            \Upsilon_{\mathbf{k}}(\tau)=\frac{\chi_{\mathbf{k}}(\tau)}{\epsilon_{\mathbf{k}}(\tau)}, \ \ \ \ \ \Phi_{\mathbf{k}}(\tau)=\pi-\arg[\Upsilon_{\mathbf{k}}(\tau)].
            \end{eqnarray}
Eq.~\eqref{IH} implies that the instantaneous energy eigenstates are the normalized occupation number states,
\begin{eqnarray}
\ket{\Psi_{mn}^{(\mathbf{k})}(\tau)} &=& \ket{m;\alpha_{\mathbf{k}}(\tau)} \ket{n;\beta_{-\mathbf{k}}(\tau)}\nonumber\\
&=&\frac{[\beta_{-\mathbf{k}}^{\dagger}(\tau)]^{n}[\alpha_{\mathbf{k}}^{\dagger}(\tau)]^{m}\ket{0; \alpha_{\mathbf{k}}(\tau)}\ket{0; \beta_{-\mathbf{k}}(\tau)}}{\sqrt{m!n!}},\nonumber\\
\label{SQEESt}
\end{eqnarray}
at each instant $\tau$.
$\ket{\Psi_{00}^{(\mathbf{k})}(\tau)}=\ket{0; \alpha_{\mathbf{k}}(\tau)}\ket{0;\beta_{-\mathbf{k}}(\tau)}$ is the corresponding instantaneous two-mode vacuum state.

Following Eqs.~\eqref{eq:BTfinal} and \eqref{SQEESt}, we determine the instantaneous energy eigenstates in the $(\alpha, \beta)$-mode after the perturbation to be 
\begin{eqnarray}
\ket{\Psi_{mn}^{(\mathbf{k})}(\tau)} = \sum_{l=0}^\infty(-1)^{l}P_{mn}^{(\mathbf{k})}(\tau, l)
\ket{l+\delta_{\alpha}; \alpha_{\mathbf{k}}} \ket{l+\delta_{\beta}; \beta_{-\mathbf{k}}},\nonumber\\
\label{EES}
\end{eqnarray}
where $(\delta_{\alpha}, \delta_{\beta})=(\frac{\delta+|\delta|}{2}, \frac{|\delta|-\delta}{2})$, $\delta=m-n$, and
\begin{eqnarray}
P_{mn}^{(\mathbf{k})}(\tau, l)=\frac{1}{\sqrt{m!n!}}
\left(\frac{1}{u_{\mathbf{k}}}\right)^{|\delta|}\left(\frac{1}{u_{\mathbf{k}}v_{\mathbf{k}}}\right)^{\mu}q^{(\mu, |\delta|)}_{l; \mathbf{k}}P_{00}^{(\mathbf{k})}(\tau, l)\nonumber\\ 
\label{PRRR}
\end{eqnarray}
with $u_{\mathbf{k}}=u_{\mathbf{k}}(\tau)$, $v_{\mathbf{k}}=v_{\mathbf{k}}(\tau)$, $\mu = \min\{m, n\}$, and 
\begin{eqnarray}
P_{00}^{(\mathbf{k})}(\tau, l)=\left[-\frac{v_{\mathbf{k}}(\tau)}{u_{\mathbf{k}}(\tau)}\right]^{l}=\frac{\left[-e^{i\Phi_{\mathbf{k}}(\tau)}\tanh\Theta_{\mathbf{k}}(\tau)\right]^{l}}{\cosh \Theta_{\mathbf{k}}(\tau)}.\ \ \ 
\end{eqnarray}
The coefficients $q^{(\mu, |\delta|)}_{l; \mathbf{k}}$ are given by the following recursive relations
\begin{eqnarray}
q^{(\mu, |\delta|>0)}_{l; \mathbf{k}}&=&|u_{\mathbf{k}}|^{2}\sqrt{l+|\delta|}q^{(\mu, |\delta|-1)}_{l; \mathbf{k}}
-|v_{\mathbf{k}}|^{2}\sqrt{l+1}q^{(\mu, |\delta|-1)}_{l+1; \mathbf{k}}, \nonumber\\
q^{(\mu>0, 0)}_{l; \mathbf{k}}&=&l|u_{\mathbf{k}}|^{4}q^{(\mu-1, 0)}_{n-1; \mathbf{k}}-(2l+1)|u_{\mathbf{k}}v_{\mathbf{k}}|^{2}q^{(\mu-1,0)}_{l; \mathbf{k}}\nonumber\\
&&+(l+1)|v_{\mathbf{k}}|^{4}q^{(\mu-1,0)}_{l+1; \mathbf{k}},
\label{RRRP}
\end{eqnarray}
such that $q^{(0, 0)}_{l; \mathbf{k}}=1$ for each $l$.

We note that the energy eigenstates of the system become entangled in the $(\alpha, \beta)$-mode after the 
light-induced perturbation, taking the form of two-mode entangled squeezed states, as given in Eqn.\eqref{EES}. In contrast, the energy eigenstates before the perturbation, given in Eqn.\eqref{SQEES}, are product states in the $(\alpha, \beta)$-mode. This entanglement can be quantified using the entropy of entanglement, defined as
\begin{eqnarray}
E\left[\ket{\Psi_{mn}^{(\mathbf{k})}(\tau)}\right] &=& -\sum_{l=0}^{\infty} |P_{mn}^{(\mathbf{k})}(\tau, l)|^{2} \log |P_{mn}^{(\mathbf{k})}(\tau, l)|^{2}, \quad \quad
\label{EIE}
\end{eqnarray}
or, equivalently, by the Schmidt rank, defined as the number of nonzero elements in Eq.~\eqref{EES}, i.e., the cardinality $\#\{|P_{mn}^{(\mathbf{k})}(\tau, l)|^{2} \ne 0 \}_{l=0}^{\infty}$. A Schmidt rank of one indicates that the state is a product state, while a rank greater than one implies that the state is entangled.

\begin{figure}[t]
\begin{center}
\includegraphics[width=80mm]{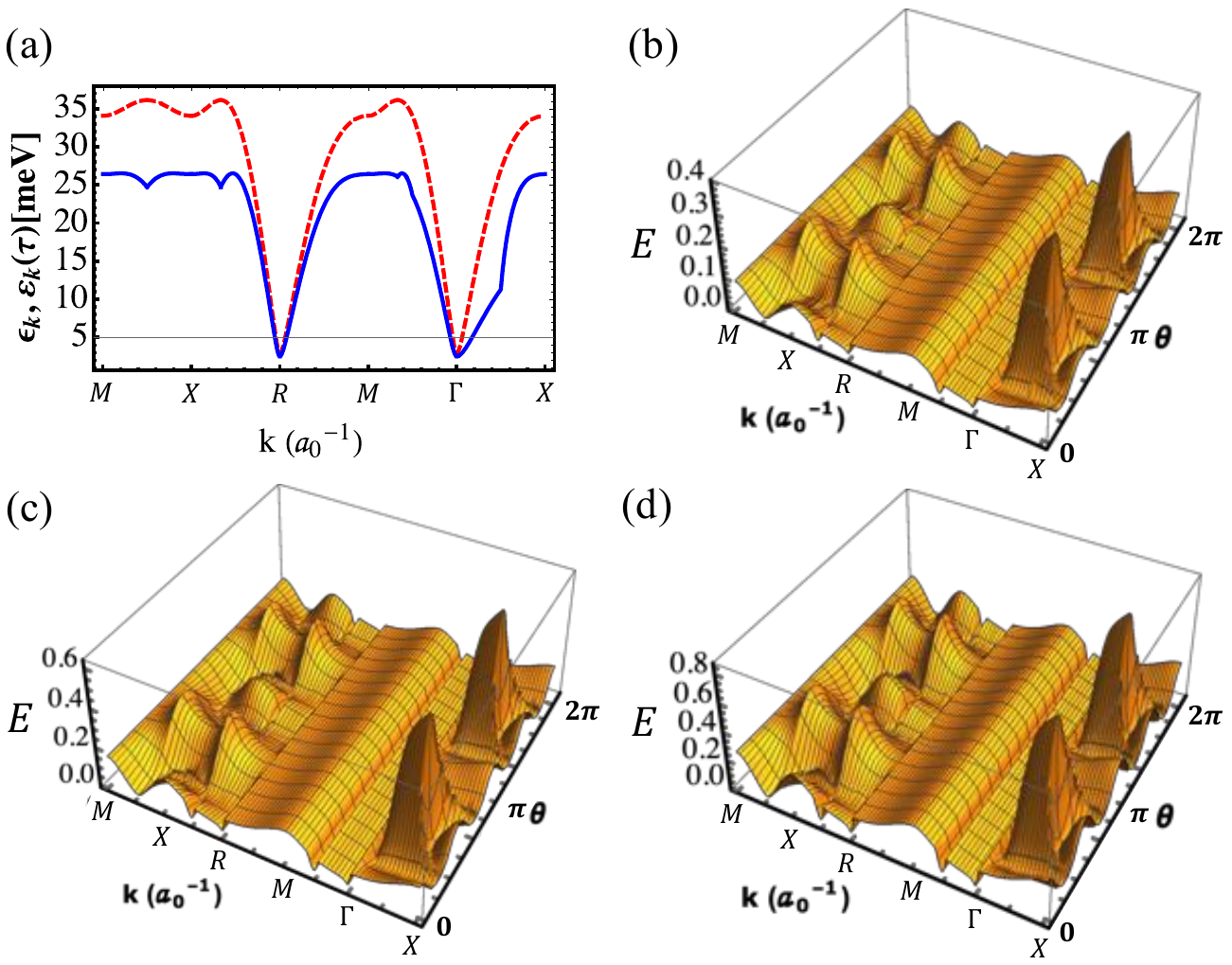}
\end{center}
\caption{(Color online) Plot~(a) illustrates the dispersion relation before (dashed red) and after (solid blue) the ultrafast light-matter interaction. Shake-up generated magnon-magnon entanglement, $E$, between $\alpha$ and $\beta$ modes that reside in instantaneous energy eigenstates are showm in:   
(b)~$\ket{\Psi_{00}^{(\mathbf{k})}(\tau)}$,  
(c)~$\ket{\Psi_{10}^{(\mathbf{k})}(\tau)} \equiv \ket{\Psi_{10}^{(\mathbf{k})}(\tau)}$, and  (d)~$\ket{\Psi_{11}^{(\mathbf{k})}(\tau)}$.  
$E$ is plotted as a function of the in-plane electric field angle $\theta$, along a symmetric path in the first Brillouin zone. The system is modeled using a simple cubic spin-1/2 lattice, with the pulse envelope in Eq.~\eqref{eq:pulse} set to $f(\tau) = 1$. The entanglement landscape is identical for different reference magnon states, but its strength increases with the number of magnons.}
\label{fig2}
\end{figure}

We present in Fig.\ \ref{fig2} examples of magnon-magnon entanglement entropy, which will be used to illustrate the connection between shake-up and entanglement. Figure \ref{fig2}$(a)$ shows the magnon dispersion of the original, unperturbed Hamiltonian specified in Eq.\ \eqref{BSM}, alongside the dispersion resulting from the modified Hamiltonian in Eq.\ \eqref{TDTH}. The entanglement, evaluated immediately after the excitation-induced modification of the Hamiltonian in Eq.\ \eqref{TDTH}, are shown in Fig.\ \ref{fig2}$(b)$–$(d)$. We consider here an antiferromagnetic simple cubic spin-$\frac{1}{2}$ lattice subjected to an in-plane laser field, ${\bf E} = E(\cos\theta, \sin\theta, 0)$. The parameters used in the calculation are: $J = 12$ meV, $\mathcal{K} = 0.01 J$, $\Delta J = -\frac{8}{10} J$, and $B = 0$. In all three cases, the strongest excitation-driven entanglement appears at $\mathbf{k}$-points located halfway between the $\Gamma$ and $X$ points when the electric field is aligned with the momentum vector $\mathbf{k}$. This behavior results from the maximal projection of the electric field onto the nearest-neighbor bonds of the lattice (see Eq.\ \eqref{eq:DeltaJ}).

If the system is initially in one of the eigenstates of the original Hamiltonian in Eq.\ \eqref{BSM}, for instance, $\ket{\psi_{N}} = \ket{\psi_{mn}^{(\mathbf{k})}}$ with $N = n + m$, and the perturbation is abrupt, the initial state does not have time to evolve during a fast excitation process. Similar to the situation in electron spectroscopy, this gives rise to shake-up structures. The corresponding magnon shake-up structure can be observed through the associated energy fluctuations
\begin{eqnarray}
\bra{\psi_{mn}^{(\mathbf{k})}} H_{\mathbf{k}}(\tau)\ket{\psi_{mn}^{(\mathbf{k})}}
&=&|P_{mn}^{(\mathbf{k})}(\tau, \mu)|^{2}[2\mu\varepsilon_{\mathbf{k}}(\tau)+\delta B]+\nonumber\\
&&\sum_{l=0, l\ne\mu}^{\infty}|P_{mn}^{(\mathbf{k})}(\tau, l)|^{2}[2l\varepsilon_{\mathbf{k}}(\tau)+\delta B],\nonumber\\ 
\label{EFSHUP}
\end{eqnarray}
which follows from the shake-up state 
\begin{eqnarray}
\ket{\psi_{mn}^{(\mathbf{k})}} &=&P_{mn}^{(\mathbf{k})}(\tau, \mu)\ket{\mu+\delta_{\alpha}; \alpha_{\mathbf{k}}(\tau)} \ket{\mu+\delta_{\beta}; \beta_{-\mathbf{k}}(\tau)} 
\nonumber\\
& + & \sum_{l=0, l\ne\mu}^{\infty}P_{mn}^{(\mathbf{k})}(\tau, l)\ket{l+\delta_{\alpha}; \alpha_{\mathbf{k}}(\tau)} \ket{l+\delta_{\beta}; \beta_{-\mathbf{k}}(\tau)}\nonumber\\
&& = \ket{\psi_{mn; R}^{(\mathbf{k})}}+\ket{\psi_{mn: E}^{(\mathbf{k})}},
\end{eqnarray}
obtained through Eqs. \eqref{SQEES} and \eqref{eq:BTfinal}. 
\begin{figure}[t]
\begin{center}
\includegraphics[width=64mm]{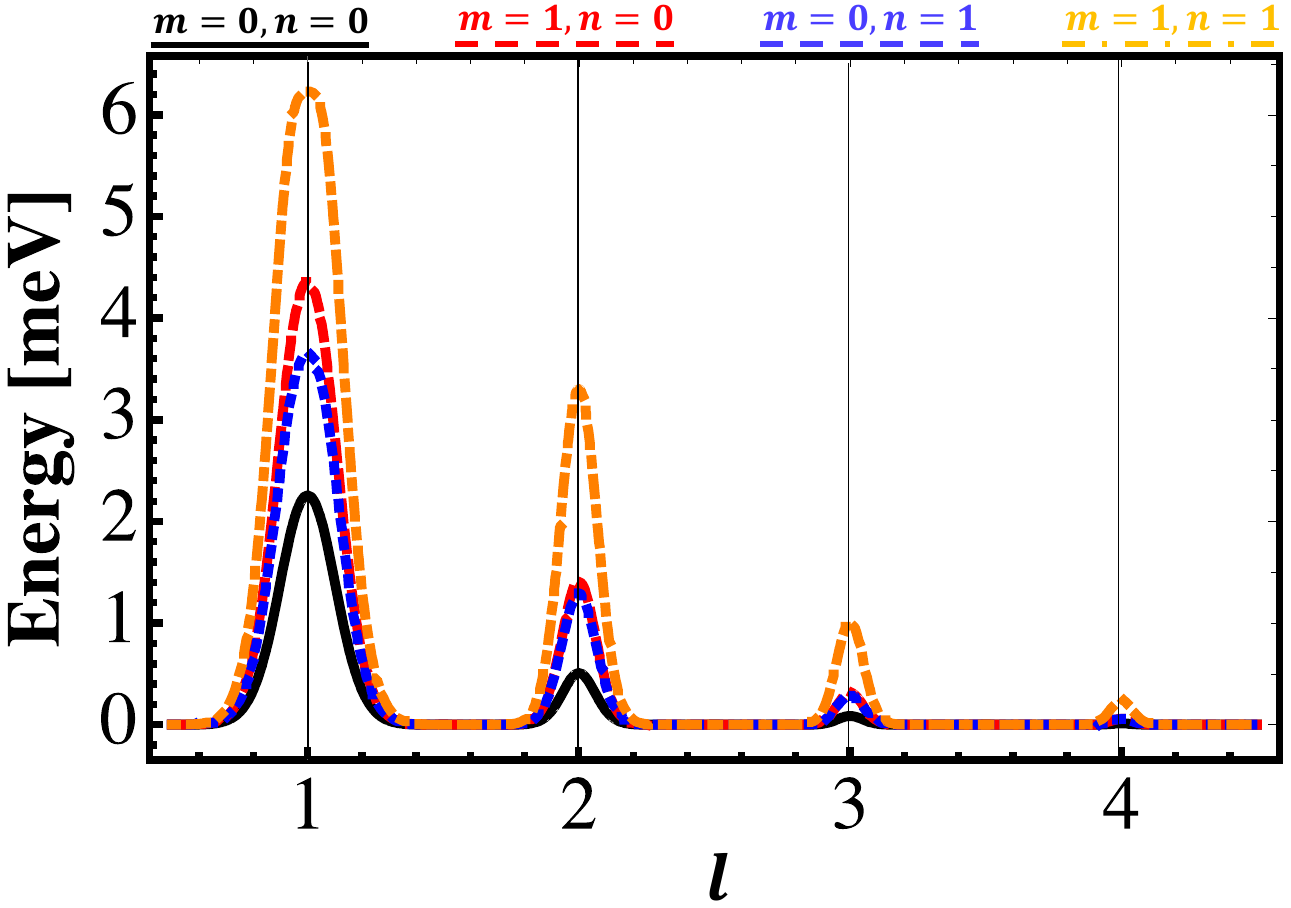}
\end{center}
\caption{(Color online). The first four shake-up peaks, which correspond to the points of maximum entanglement in Fig.\ \ref{fig2}, at $\mathbf{k} = (0, \pi/2, 0)$ in the Brillouin zone when the electric field is aligned with the momentum vector $\mathbf{k}$. }
\label{fig:Shake-up}
\end{figure}

The second term on the right-hand side of Eq.~\eqref{EFSHUP} represents the total energy of the magnon shake-up structure. For reference magnon numbers $n$ and $m$, this energy is obtained by summing contributions from sideband magnon states labeled by harmonic index $l$, $[2l\varepsilon_{\mathbf{k}}(\tau)+\delta B]$, each weighted by the transition probability $|P_{mn}^{(\mathbf{k})}(\tau, l)|^{2}$ between the reference and $l$-th occupation magnon states. These contributions appear as shake-up peaks in the energy fluctuation spectrum following an abrupt perturbation of the magnetic system.
Figure \ref{fig:Shake-up} shows the magnon shake-up structures evaluated directly after excitation, from regions of maximum entanglement in Fig.~\ref{fig2}. The system, modeled with the same parameters as in Fig.~\ref{fig2}, is driven by a Gaussian-like pulse, $f(t) = \exp[-20 (t - t_0)^2]$, with a time shift $t_0 \propto l$, as used in Eq.~\eqref{EFSHUP}. The figure displays the first four shake-up peaks for various reference magnon states. While lower $l$ values contribute less energy, they dominate due to the transition probability decreasing as a function of $l$.

We end our discussion by a theorem that establishes a concrete connection between magnon entanglement and the shake-up structure. 

 {\it Theorem:}  Magnon shake-up structures serve as conclusive evidence of entanglement between magnon modes $\alpha$ and $\beta$. 
 
{\it Proof:} It is evident that $\#\{|P_{mn}^{(\mathbf{k})}(\tau, l)|^{2} \ne 0\}_{l=0}^{\infty} > 1$ if $\#\{|P_{mn}^{(\mathbf{k})}(\tau, l)|^{2} \ne 0\}_{l=0,, l \ne \mu}^{\infty} > 0$. The former indicates a Schmidt rank greater than one, thereby confirming entanglement in the instantaneous eigenstates of Eq.\ \eqref{EES} in the $(\alpha, \beta)$ modes. The latter confirms the presence of a nonzero shake-up structure in the associated energy fluctuations described by Eq.\ \eqref{EFSHUP}.
This proves that the observation of a magnon shake-up structure confirms the presence of entanglement between magnon modes $\alpha$ and $\beta$ residing in the instantaneous energy eigenstates.

In summary, we have demonstrated shake-up structures in a quantum description of magnons, within the sudden approximation. Our findings establish magnon shake-up features as clear signatures of entanglement between magnon modes. While our primary focus has been on antiferromagnetic systems, where quantum effects are pronounced \cite{azimi-mousolou2020, azimi-mousolou2021, azimi-mousolou2023, Liu2023}, we have also observed qualitatively similar phenomena in ferromagnetic systems, where the two magnon modes are distinguished by transverse anisotropy. These results provide a solid foundation for further investigation of quantum correlations in spin-based platforms and open promising avenues for quantum information applications in magnonics.

\section{Acknowledgement}
A.D. and O.E. acknowledge support from the Wallenberg Initiative Materials Science for Sustainability (WISE) funded by the Knut and Alice Wallenberg Foundation (KAW). O.E. also acknowledges support from STandUPP, eSSENCE, the Swedish Research Council (VR), the European Research Council through the ERC Synergy Grant 854843-FASTCORR and the Knut and Alice Wallenberg Foundation (KAW-Scholar program) and NL-ECO: Netherlands Initiative for Energy-Efficient Computing (with project number NWA. 1389.20.140) of the NWA research program. 
Financial support from the
Swedish Research Council (Vetenskapsrådet, VR) Grant No. 2016-05980, Grant No. 2019-05304, and Grant No. 2024-04986, and the Knut and Alice Wallenberg foundation Grant No. 2018.0060, Grant No. 2021.0246, and Grant No. 2022.0108 is acknowledged.  
The computations/data handling were enabled by resources provided by the National Academic Infrastructure for Supercomputing in Sweden (NAISS), partially funded by the Swedish Research Council through grant agreement no. 2022-06725.
Valuable discussions with Prof.\,Olof Karis, Dr.\,Tom Silva, Dr.\,Oscar Grånäs, Prof. Hermann D\"urr and Dr.\,Mohamed Elhanoty are acknowledged.


\begin{thebibliography}{}
\bibitem{note} The total hamiltonian is of the form $H=\sum_{\mathbf{k}}H_{\mathbf{k}}$, where $H_{\mathbf{k}}$ for different $\mathbf{k}$s are decoupled.
\end{thebibliography}
            \end{document}